\documentclass[aps,prl,twocolumn,floatfix,superscriptaddress,showpacs,nofootinbib]{revtex4}

\usepackage{graphicx}     
\usepackage{dcolumn}      
\usepackage{bm}           
\usepackage{amsmath}
\usepackage{graphics}
\usepackage{amssymb}
\usepackage{amscd}
\usepackage{afterpage}
\usepackage{float,times}
\usepackage{epsfig}
\usepackage{theorem}
\usepackage{wrapfig}
\usepackage{picinpar}

\usepackage[normalem]{ulem} 
\usepackage{color} 

\newcommand{\ket}[1]{|#1 \rangle}

\newcommand{\qed}{\nobreak \ifvmode \relax \else
      \ifdim\lastskip<1.5em \hskip-\lastskip
      \hskip1.5em plus0em minus0.5em \fi \nobreak
      \vrule height0.75em width0.5em depth0.25em\fi}

\DeclareSymbolFont{AMSb}{U}{msb}{m}{n}
\DeclareMathSymbol{\N}{\mathbin}{AMSb}{"4E}
\DeclareMathSymbol{\Z}{\mathbin}{AMSb}{"5A}
\DeclareMathSymbol{\R}{\mathbin}{AMSb}{"52}
\DeclareMathSymbol{\Q}{\mathbin}{AMSb}{"51}
\DeclareMathSymbol{\I}{\mathbin}{AMSb}{"49}
\DeclareMathSymbol{\C}{\mathbin}{AMSb}{"43}

\begin{document}

\title{Information free quantum bus for generating stabiliser states}

\author{Simon J. Devitt, Andrew D. Greentree, Lloyd C.L. Hollenberg}

\affiliation{
Centre for Quantum Computer Technology, School of Physics\\
University of Melbourne, Victoria 3010, Australia.}
\date{\today}

\begin{abstract}
Efficient generation of spatially delocalised entangled states is at
the heart of quantum information science.  Generally flying qubits
are proposed for long range entangling interactions, however here we
introduce a bus-mediated alternative for this task. Our scheme
permits efficient and flexible generation of deterministic two-qubit
operator measurements and has links to the important concepts of
mode-entanglement and repeat-until-success protocols.  Importantly,
unlike flying qubit protocols, our bus particle never contains
information about the individual quantum states of the particles,
hence is information-free.
\end{abstract}

\maketitle

\section{Introduction}

\indent \indent Recent research into quantum information and
computation has not only spawned a number of architecture proposals
for quantum computation (QC), but also proposals for useful
technologies based on smaller, entangled quantum systems. Quantum
key distribution \cite{QKD}, quantum dense coding \cite{QI} and the
improvement of frequency standards using entangled systems \cite{QL}
are strong examples of how highly entangled states can be exploited
to build novel devices: The incorporation of such entangled-state
protocols with existing micro- and nano-scale fabrication is an
enormous opportunity for the semiconductor industry.

The development of viable quantum computers and preparation of
effective multi-qubit entangled states depends crucially on
transport protocols that can be used to shuttle quantum information
and to allow for interactions between isolated qubits. Effective
transport of quantum information is essential to the scaling of
small, functional elements, and will enable an interpolation between
small scale devices and full-blown, massively entangled quantum
computers.  In the solid-state, inevitable fabrication errors also
enforce the need for defect-tolerant methodologies, and transport
allows for natural mechanisms to incorporate such features.
\\
\indent Ion traps \cite{ion,qccc} and photonic based quantum
computers \cite{KLM} generally allow for easy and quick long range
qubit transport.  On the other hand, solid-state architectures
\cite{ss3,ss1,ss2,Ladd2002,ss4,deSousa2004,Hill2005} are often
limited to nearest neighbour interactions on a linear array of
qubits, leading to several problems: For example, qubit transport in
linear systems is generally proposed using SWAP operations which
reduce the threshold for concatenated error correction and unless
modifications to the underlying architecture are made, such schemes
are not Fault-Tolerant \cite{LNN}.
\\
\indent The concept of flying qubits and quantum bus systems has
received significant attention \cite{fq1,fq2,CTAP,fq5,lloyd} as a
means to combat the problem of long range transport in systems that
do not exhibit them naturally.  Bose introduced quantum state
transfer via unmodulated spin chains \cite{Bose}, while
teleportation hubs \cite{Ike} have been proposed to combat long
range transport in linear systems. However, these transport hubs do
not remove the need for SWAP gates between qubits \emph{not}
adjacent to hubs and generally do not allow small multi-qubit state
preparation without considerable resource overhead. Here, we show a
transport protocol that mediates entangling operations between
isolated data qubits, without communicating single-qubit states
directly.  We show that this scheme is extremely flexible and
efficient in generating multi-qubit entangled states and operates in
a fundamentally different fashion from conventional gate-driven
entangling operations.  Furthermore, the flexibility of our scheme
relaxes some of the requirements for controllability, and hence
could be an enabler for new approaches to quantum computing.
\\
Along with flying qubit schemes and quantum bus systems,
interactions on well isolated data qubits can be achieved using
measurement based quantum computation.  The two main concepts are
teleportation based QC (TQC) \cite{TQC1,TQC2,TQC4,TQC5} and cluster
state QC (CSQC) \cite{cl1,cl3}.  These ideas differ significantly
from the traditional circuit based paradigm in that interactions are
not performed using unitary gates.  TQC uses correlated multi-qubit
measurements and appropriate ancilla states to perform  operations
via teleportation.  CSQC requires an initial, highly entangled,
multi-qubit cluster (or general graph state \cite{graph}) and
arbitrary single qubit measurements.
\\
\indent To mediate interactions between two qubits, we introduce a
bus qutrit.  As an example we show a spatial qutrit defined over
three sites, which we label Alice, Bob$_1$, and Bob$_2$,
$|A\rangle$, $|B_1\rangle$ and $|B_2\rangle$ respectively. The
qutrit takes on the role of the ancilla used in standard two-qubit
operator measurements \cite{Nielsen}. Our scheme utilises several
properties from circuit, teleportation and cluster state
computation, presenting a hybrid protocol that could be used to
achieve universality.
\\
\section{Qutrit transfer protocol}

\indent To generate operator measurements, the qutrit is placed into
a superposition of well-defined non-local spatial states, adjacent
to separate isolated data qubits. The spatial state of the qutrit is
then used as a control for unitary gates on the data qubits. This
distinguishes our protocol from traditional flying qubit schemes in
that computational information stored on each qubit is never
transferred to or by the bus, justifying the term
\emph{information-free}. Some potential implementations include an
electron that can be placed into a physically delocalized
superposition around data qubits, or a photon pulse that can be
placed into a superposition of spacio-temporal modes.

\begin{figure}[tb!]
\includegraphics[width=0.7\columnwidth,clip]{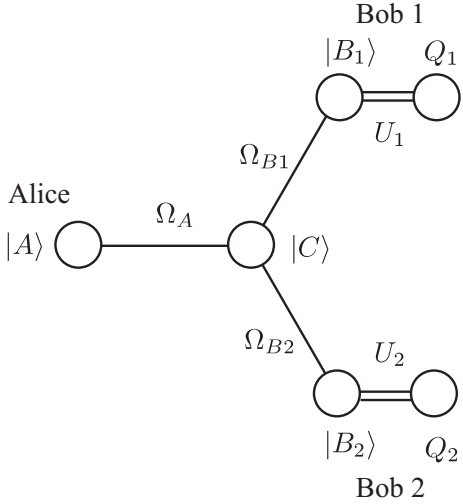}
\caption{\label{fig:MRAPdiag} Schematic of the configuration
required to demonstrate the qutrit transport protocol using a
spatially defined bus qubit.  In common with communications
approaches, we define the starting site as Alice, and the two
recipient points are Bob$_1$ and Bob$_2$. Using the multi-recipient
adiabatic passage (MRAP) protocol, the central site is never
occupied. By using counter-intuitive pulse ordering and varying the
relative intensities of the tunnelling matrix elements, it is
possible for Alice to send the bus particle to an arbitrarily
weighted superposition of the Bobs, although for our purposes the
equally weighted superposition is chosen. Here the single lines
correspond to controlled tunnelling matrix elements for the single
control particle Hamiltonian, and the double lines correspond to the
controlled interactions between the Bob sites and the qubits. The
parity result from the two-qubit operator measurement on $Q_1$ and
$Q_2$ is effected by post-selecting the state of the qutrit
following the reversal of the protocol.}
\end{figure}

For clarity, we will concentrate on spatially delocalised electrons,
using the specific protocol of multi-recipient adiabatic passage
(MRAP)~\cite{MRAP}. MRAP is a protocol for adiabatically generating
spatially delocalised superposition states appropriate for
solid-state quantum devices.  This case is illustrated in Fig.
\ref{fig:MRAPdiag}, and we follow the sender/receiver notation
commonly used in communications theory and in Ref.~\cite{MRAP}. Note
that the counter-intuitive pulse sequence used for MRAP excludes
population from the central dot (labeled $|C\rangle$) and hence we
ignore this state in our analyses that follow.

\indent We present the Qutrit Transfer Protocol (QTP) as a standard
transformation that can be applied to a given system, demonstrating
the equivalence of this scheme to measurement of the operators
$XX=\sigma_x \otimes \sigma_x$ and $ZZ=\sigma_z\otimes \sigma_z$.
Using the QTP and the ability to perform single qubit operations
directly on data qubits we demonstrate linear cluster state
preparation and universal computation.
\\
\indent The general transformations for a qutrit defined by a source
state, (Alice, $|A\rangle$) and target states (Bobs, $|B_1\rangle$,
$|B_2\rangle$) is described by the vector,
$(|A\rangle,|B_1\rangle,|B_2\rangle)^T$.  The essential
transformation matrix can be written as,
\begin{equation}
\begin{aligned}
U_{\text{QTP}} = \begin{pmatrix}
0 & 1/\sqrt{2} & 1/\sqrt{2} \\
1/\sqrt{2} & 1/2 & -1/2 \\
1/\sqrt{2} & -1/2 & 1/2
\end{pmatrix}.
\end{aligned}
\label{eq:Trans}
\end{equation}

\begin{figure}[t]
\epsfig{figure=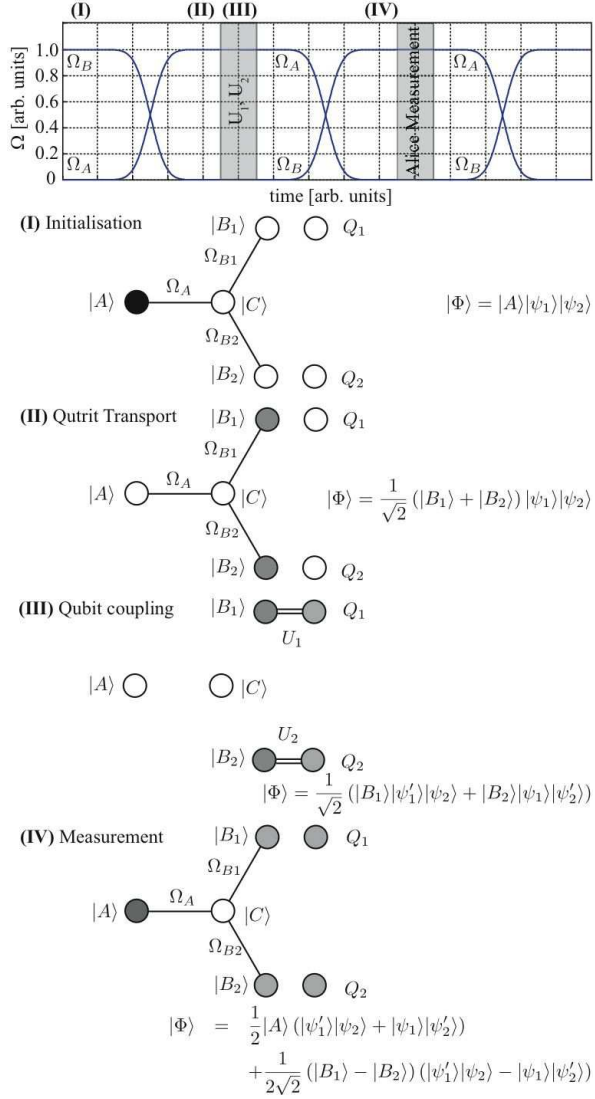,width = 0.9\columnwidth}
 \caption{Pulsing sequence for tunneling matrix elements (top)
and schematics showing system evolution through the MRAP protocol to
realise qutrit transport and two-qubit operator measurements. (I)
The qutrit is initialised at $\ket{A}$ with the two qubits in some
state $\ket{\psi} = \ket{\psi_1}\ket{\psi_2}$. (II) MRAP takes the
qutrit to the state $(\ket{B_1} + \ket{B_2})/\sqrt{2}$, (III) the
controlled unitaries ($U_1, U_2$) are performed between each Bob
site and the qubits, conditional on the presence of the qutrit at
the appropriate site, and the system is transformed to
$(\ket{B_1}\ket{\psi_1'}\ket{\psi_2} +
\ket{B_2}\ket{\psi_1}\ket{\psi_2'})/\sqrt{2}$. (IV) the transport is
reversed and a projective measurement of the qutrit at $\ket{A}$
performed. The results of this measurement projects the qubits into
an eigenstate of the operator $U_1 U_2$, and depending on the
measurement result, a phase flip at either $\ket{B_1}$ or
$|\ket{B_2}$ and further qutrit transport protocol can be used to
return the qutrit to Alice.}
 \label{fig:qutrit}
\end{figure}

As stated above, the MRAP protocol provides a natural method for
generating the required qutrit transformations, and we briefly
review these for clarity.  The form of a four-site structure as
shown in Fig.~\ref{fig:MRAPdiag} is isomorphic to the well-known
tripod atom familiar to quantum optics.  In particular, we can apply
techniques for generating arbitrary ground-state superpositions that
have already been developed in the optical regime in this case
\cite{bib:UnanyanPRA1999,bib:LukinPRL2000}.  Defining the coupling
between each site and the central dot as $\Omega_{\alpha}$ for
$\alpha = A, B_1, B_2$ (assumed real and positive), where Alice and
each Bob has control of the energy of the bus-particle on their
site, which has been assumed equal, and their appropriate tunnel
matrix element, we have the Hamiltonian
\begin{multline}
\mathcal{H} = \Omega_A (t) |C\rangle\langle A| + \Omega_{B1} (t)
|C\rangle \langle B_1| + \Omega_{B2} (t) |C\rangle \langle B_2| +
\mathrm{h.c.}
\end{multline}
where the time-varying tunneling matrix elements are controlled, for
example, by local control of surface gate potentials.

The relevant states for MRAP state transfer are those with zero
energy eigenvalue, which are given by the null-space of
$\mathcal{H}$.  These are
\begin{eqnarray}
|D_1\rangle = \frac{\Omega_{B1}}{\sqrt{\Omega_A^2 + \Omega_{B1}^2}}
|A\rangle - \frac{\Omega_{A}}{\sqrt{\Omega_A^2 + \Omega_{B1}^2}}
|B_1\rangle, \\
|D_2\rangle = \frac{\Omega_{B2}}{\sqrt{\Omega_A^2 + \Omega_{B2}^2}}
|A\rangle - \frac{\Omega_{A}}{\sqrt{\Omega_A^2 + \Omega_{B2}^2}}
|B_2\rangle,
\end{eqnarray}
where we have dropped the time dependence of the $\Omega$. Of
course, any superposition of these vectors is also in the null
space, so if the system remains in the null-space, $|C\rangle$ will
never be populated, hence the definition of the bus particle in this
case as a qutrit defined over sites $|A\rangle$, $|B_1\rangle$, and
$|B_2\rangle$. In particular, we set $\Omega_{B1} = \Omega_{B2} =
\Omega_B = \Omega_B^{\max} [1 - \mathrm{erf}(t/\sigma)]/2$ and
$\Omega_A = \Omega_A^{\max} [1+\mathrm{erf} (t/\sigma)]/2$, where
$\sigma$ is the width of the roll off of the error function (erf),
$\Omega_A^{\max}$ and $\Omega_B^{\max}$ are the maximum values of
the tunneling matrix elements.  Without presenting details here, we
simply note that either $\sigma$ or the $\Omega^{\max}$ should be
chosen to maintain adiabaticity, then the state that adiabatically
connects Alice to each of the Bobs is
\begin{eqnarray}
|D_3\rangle = \frac{2\Omega_B |A\rangle - \Omega_A (|B_1\rangle +
|B_2\rangle)}{\sqrt{4 \Omega_B^2 + 2 \Omega_A^2}}.
\end{eqnarray}

\section{Two-qubit operator measurements}

\indent Consider two data qubits $[Q_1,Q_2]$ that can be coupled to
the qutrit through a controlled two-particle entangling operation.
For example, we could reexpress this interaction as either a CNOT or
CZ gate where the control parameter is the presence or absence of
the wavefunction of the qutrit at the nearest physical location.
Such an interaction is physically motivated as, for example, we
could consider a Coulomb interaction providing the entangling
operation, and hence if the particle is not present, no interaction
will occur. The controlled gate is applied to $Q_1$ ($Q_2$)
\emph{iff} the qutrit state has non-zero amplitudes for
$|B_1\rangle$ ($|B_2\rangle$) respectively. Alice transmits the
particle to an equal superposition of the Bob sites and then the
particle is coupled to the data qubits ($[Q_1, Q_2]$) through a CNOT
(or CZ) operation. The protocol is schematically represented in
Fig.~\ref{fig:qutrit}.  For total system state $\ket{\Phi}$ and a
general two qubit state $|\psi\rangle_{Q_1,Q_2}$, these
transformations are
\begin{equation}
\begin{aligned}
U_{\text{QTP}}|\Phi\rangle = &U_{QTP}|A\rangle\otimes|\psi\rangle
=\frac{1}{\sqrt{2}}(|B_1\rangle+|B_2\rangle)\otimes|\psi\rangle \\
&\overset{CNOT}{\Longrightarrow} \frac{1}{\sqrt{2}}(|B_1\rangle
X_{Q_1}|\psi\rangle+|B_2\rangle X_{Q_2}|\psi\rangle),
\end{aligned}
\end{equation}
where $X_{Q_i}|\psi\rangle$ is a bit flip on qubit $Q_i$.
The QTP is performed again, transforming the state to,
\begin{equation}
\begin{aligned}
&\frac{1}{2}|A\rangle(X_{Q_1}|\psi\rangle+X_{Q_2}|\psi\rangle) + \\
&\frac{1}{2\sqrt{2}}(|B_1\rangle -
|B_2\rangle)\otimes(X_{Q_1}|\psi\rangle-X_{Q_2}|\psi\rangle).
\end{aligned}
\end{equation}
The system is then measured to determine if the qutrit has returned
to the state $|A\rangle$. If it has, the information qubits are
projected to $(X_{Q_1}|\psi\rangle+X_{Q_2}|\psi\rangle)$. If not,
the information qubits are projected to
$(X_{Q_1}|\psi\rangle-X_{Q_2}|\psi\rangle)$.  Irrespective of the
result after measurement, the transport qutrit is completely
decoupled from the data qubits, therefore it can be discarded.
However, if it needs to be reused and was not measured to the Alice
state, a phase flip can be applied to $B_2$ taking,
$(|B_1\rangle_{1}-|B_2\rangle_{1})\otimes(X_{Q_1}|\psi\rangle-X_{Q_2}|\psi\rangle)
\rightarrow
(|B_1\rangle_{1}+|B_2\rangle_{1})\otimes(X_{Q_1}|\psi\rangle-X_{Q_2}|\psi\rangle)$.
After the phase flip the protocol can be deterministically reversed
and the qutrit will return to the $|A\rangle$ state for reuse.
\\
\indent Inspection of the states generated by the protocol above,
$(X_{Q_1}|\psi\rangle+X_{Q_2}|\psi\rangle)$ and
$(X_{Q_1}|\psi\rangle-X_{Q_2}|\psi\rangle)$, reveals that if the
qutrit is measured at Alice, the data qubits are projected to a $+1$
eigenstate of $X_{Q_1}X_{Q_2}$. If the qutrit is not measured at
Alice then the data qubits are projected to a $-1$ eigenstate of
$X_{Q_1}X_{Q_2}$.  The same analysis can be performed using a CZ
interaction between the qutrit and data qubits. In this case, the
resultant state after measurement is projected to a $\pm 1$
eigenstate of $Z_{Q_1}Z_{Q_2}$, dependant on whether the qutrit is
measured in the Alice state.
\\
\indent As the protocol projects the data qubits into a $\pm 1$
eigenstate of either $XX$ or $ZZ$ operators, loss of the qutrit
during the protocol does not result in data loss: At most, the loss
of the bus qutrit will simply induce a coherent $X$ or $Z$ error on
one of the qubits as appropriate.  A single qubit error is
completely contained within the qubit space, and hence can be
corrected via standard error correction protocols.  Therefore,
qutrit loss results in not knowing if the data qubits are in a $+1$
or $-1$ eigenstate of $XX$ ($ZZ$).  If the qutrit is lost,
subsequent protocols will project to the data qubits to the same
eigenstate. This leads to a repeat until success scheme
\cite{LBK,BK,LBBKK} with no additional loss protocols required. For
a practical device, therefore, it will be necessary to know the
dephasing times appropriate for the qutrit, however again, we note,
that this scheme is still considerably more robust than a
conventional flying qubit responding to an equally dephasing
environment.

It is interesting to note, that the scheme as presented has much in
common with the concept of mode entanglement
\cite{Tan1991,Hardy1994}. This topic has been the subject of much
discussion recently due to the apparent contradiction between
particle superposition and entanglement of modes that can be found
in even simple beamsplitting experiments.  A cursory glance of the
qutrit transfer protocol above, shows that its action is analogous
to the action of a beamsplitter on a single photon.  One can
therefore see that the reversals of the protocol provide a mechanism
for the global measurement mentioned by Ashab \textit{et al.}
\cite{Maruyama2007} in the context of massive particle
mode-entanglement.
\\
\indent The QTP has identical properties to standard two qubit
operator measurements \cite{Nielsen}.  However, instead of the usual
ancilla qubit, which is always spatially localised, and interacts
via (non-parallel) sequences of entangling gates, in our case, we
employ a qutrit placed in a spatial superposition. In this
superposition, each term acts as a control bit on \emph{separate}
spatially isolated data qubits. This couples $[Q_1,Q_2]$ via a
qutrit bus which acts to mediate the entanglement. Alternatively, we
can contrast the schemes as follows: the conventional circuit
approach uses the computational state of the ancilla to tag certain
data states, the QTP realises this tagging through the nonlocal
degrees of freedom, essentially providing distributed entanglement
in a natural fashion. The bus itself never carries any local data
from either qubit, and after measurement becomes completely
decoupled from the qubits: i.e. the bus is always information free.
As the QTP can realise measurements of the operators $XX$ and $ZZ$,
the preparation of $N$ qubit linear cluster states and universal
computation can be achieved using these operator measurements and
local unitaries on data qubits, as we will now show.
\\
\section{Generating cluster and stabiliser states}

\begin{figure}[t]
\epsfig{figure=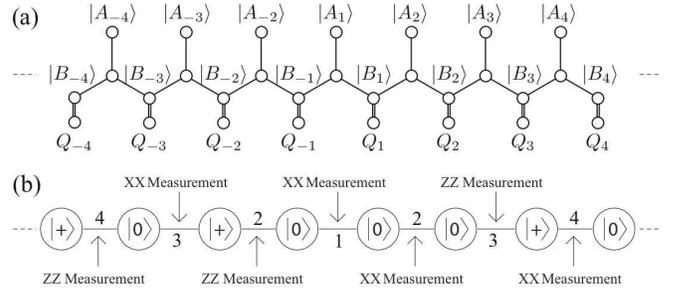,width = 1.0\columnwidth}
 \caption{(a) Schematic of multi-Alice multi-Bob structure
for generating stabilised states, the Alices are labelled so that at
time step 1, Alice$_1$ is used, and at all subsequent time steps
$i$, Alice$_i$ and Alice$_{-i}$ are used, with coupling to Bobs as
indicated. (b) Required two qubit operator measurements to prepare
an $N$ qubit linear cluster state. Each of the qubits illustrated
(circles) is initialised in the $|0\rangle$ state, Hadamard
rotations performed on specific qubits (rotating to the $|+\rangle =
(|0\rangle+|1\rangle)/\sqrt{2}$ state) and $XX$ or $ZZ$ operator
measurements made.  Each link is labelled by the respective time
step in which it can be performed, assuming that independent
operator measurements can be done in parallel by defining an
isolated $(\ket{A}, \ket{B_1}, \ket{B_2})$ channel in the bus for
each operator measurement.}
 \label{fig:linear}
\end{figure}

\indent The concept of operator measurements is closely related to
the Stabiliser formalism of Gottesman \cite{gott}, commonly used in
Quantum Error Correction (QEC) \cite{QEC1,QEC2,QEC3}.  A state
$|\Psi\rangle$ is \emph{stabilised} by a operator $U$, if
$U|\Psi\rangle = |\Psi\rangle$. For arbitrary operators, the
stabiliser formalism is generally not useful, however there exists a
certain class of states for which stabilisers provide a very elegant
analytical tool. The Clifford group, $\mathcal{C}$, is a set of
unitary operators, $\{O_j\} \in \mathcal{C}$, that under
conjugation, map elements of the Pauli group, $P_j \in \mathcal{P}$,
to themselves, $O_j^{\dagger}P_{j}O_j \in \mathcal{P}, \quad \forall
\quad [O_j,P_{j}]$. A basis set for the Clifford group consists of
CNOT, Hadamard, and $S$ gate ($S \equiv \text{diag}\{1,i\}$).  A
general $N$ qubit \emph{stabilised} state, $|\Psi\rangle_N$, can be
prepared by applying Clifford group operations to an initial
$|00...00\rangle_N$ state.  Unlike arbitrary states, stabilised
states can be described by $N$ stabilisers, $\{G_j\}_N$, instead of
the $2^N$ possible basis vectors.  A stabilised state is therefore a
simultaneous $+1$ eigenstate of each operator in $\{G_j\}_N$, which
form an abelian subgroup of the $N$ qubit Pauli group, i.e.
$\{G_j\}_N \in \{I,X,Y,Z\}^{\otimes N}$. As stabilised states can be
described by the $N$ operators $\{G_j\}_N$, quantum circuits
containing only Clifford group operations can be efficiently
simulated classically \cite{Gottesman1998}.  Universality can be
achieved by combining Clifford group operations with \emph{any}
single qubit gate that generates irrational rotations on the Bloch
sphere \cite{Nielsen}.
\\
\indent Using the stabiliser formalism, highly entangled multi-qubit
states, specifically GHZ and linear cluster states can be prepared
when only small subset of two-qubit operator measurements and single
qubit gates are available. The method requires the ability to
initialise qubits in the $|0\rangle$ state, apply single qubit
Hadamard, $X$ and $Z$ gates and the ability to perform two qubit
$XX$ and $ZZ$ operator measurements. Combining the QTP introduced
with the ability to do local operations directly on data qubits
satisfies these conditions.  GHZ preparation has already been
considered \cite{MRAP}, and the same methods can be extended to
cluster states.
\\
\indent Raussendorf and Briegel \cite{cl1,cl3} demonstrated that any
$i$-qubit cluster state, $|\text{CS}\rangle_i$, is defined by the
eigenvalue equation, $K^{(a)}|\text{CS}\rangle_i =
|\text{CS}\rangle_i$ where,
\begin{equation}
K^{(a)} = X_a \bigotimes_{b\in \text{ngbh}(a)} Z_b \quad \forall
\quad a=1 ... i, \label{eq:eq1}
\end{equation}
and ngbh($a$) represents qubits linked to site $a$ in the cluster
(neighbours), in arbitrary dimensions. Linear cluster states for 2
and 3 qubits are equivalent to Bell and three qubit GHZ states (up
to local operations).  For $N > 3$ the number of basis terms for
cluster states grow quickly, hence it is better to express large
cluster states via the eigenvalue equations. For example, a 4 qubit
linear cluster state can be generated by the operators,
\begin{equation}
\begin{aligned}
K^{(1)} &= XZII, \quad \quad
K^{(2)} &= ZXZI, \\
K^{(3)} &= IZXZ, \quad \quad
K^{(4)} &= IIZX,
\end{aligned}
\label{eq:e3}
\end{equation}
where the $\otimes$ signs are omitted for notational convenience.
Since a four-qubit cluster state satisfies,
$K^{(a)}|\text{CS}\rangle_4 = |\text{CS}\rangle_4$, the operators
$K^{(a)}$ form a basis set of the stabiliser group for a 4 qubit
linear cluster state.  The stabiliser group can be used to specify
the topology of a given cluster state, without having to write out
the state directly.
\\
\indent Linear cluster state preparation using the QTP can be
achieved by examining the stabiliser structure.  The stabilisers for
$N$ qubits are generated by (neglecting identity operators),
$K^{(1)} = X_1Z_2$, $K^{(N)} = Z_{N-1}X_N$ and $K^{(j)} =
Z_{j-1}X_jZ_{j+1}$, where $j=[2,3,...,N-1]$. Preparing a state that
satisfies this stabiliser structure using only $XX$ and $ZZ$
operator measurements, combined with direct single qubit operations
requires linking the cluster together sequentially.  To show the
method explicitly, we detail the required steps needed to prepare a
4 qubit linear cluster state, after which adding links and expanding
the cluster is straight forward.  The analysis to follow assumes
that we always measure the $+1$ eigenstate of any given operator
(i.e. the qutrit is measured at Alice), if the $-1$ eigenstate is
obtained, simply apply local $X$ and/or $Z$ gates to correct.
However, since $X$ and $Z$ gates are part of the Clifford group, all
these corrections can be applied at the end of the state
preparation.
\\
\indent Begin by initialising four qubits in the state $|\phi\rangle
= |0\rangle_{Q_1}|0\rangle_{Q_2}|0\rangle_{Q_3}|0\rangle_{Q_4}$. The
stabiliser group can be generated by the 4 operators, $Z_j$,
$j=[1,2,3,4]$. Measuring the operator $IXXI$, via the QTP, will
project the state into a stabilised eigenstate of $IXXI$ and remove
all existing stabilisers that anti-commute with $IXXI$.  In this
case, the stabilisers $IZII$ and $IIZI$ are removed, while $IZZI$
commutes with $IXXI$ and hence remains in the group. After
measurement, the state of the computer will be stabilised by the
basis operators, $K^{(1)} = IXXI$, $K^{(2)} = IZZI$, $K^{(3)} =
IIIZ$ and $K^{(4)} = ZIII$. \indent Qubits 2 and 3 are now in an
entangled Bell state described by the basis stabilisers $IXXI$ and
$IZZI$, while qubits 1 and 4 remain un-entangled.  We now perform
single qubit Hadamard rotation on qubit 1 which transforms $X$
operators to $Z$ and visa versa. This transforms the above basis
stabilisers to, $K^{(1)} = IXXI$, $K^{(2)} = IZZI$, $K^{(3)} = IIIZ$
and $K^{(4)} = XIII$. Combining the stabilisers $[IXXI, XIII]$ and
$[IZZI,IIIZ]$ shows that the qubits are also stabilised by the
operators $[XXXI,IZZZ]$.
\\
\indent To produce the 4 qubit linear cluster state now requires
$ZZII$ and $IIXX$ operator measurements. As these operator
measurements act on independent qubits they can be performed in
parallel given independent Alices. This will project the qubits into
an eigenstate of these stabilisers and remove all previous
non-commuting operators.  From the above stabilisers, $K^{(1)} =
IXXI$, $K^{(2)} = IZZI$, $K^{(3)}=IIIZ$ and $K^{(4)}=XIII$
anti-commute with either $ZZII$ or $IIXX$, while the stabilisers
$XXXI$ and $IZZZ$ commute and hence remain in the set. The qubit
register will now be in the stabilised state generated by the
operators, $K^{(1)} = ZZII$, $K^{(2)} = XXXI$, $K^{(3)} = IZZZ$ and
$K^{(4)} = IIXX$. If a Hadamard rotation is performed on qubits 1
and 3 the stabiliser group is rotated to,
\begin{equation}
\begin{aligned}
K^{(1)} &= XZII, \quad \quad
K^{(2)} &= ZXZI, \\
K^{(3)} &= IZXZ, \quad \quad
K^{(4)} &= IIZX,
\end{aligned}
\end{equation}
which is the basis set of stabiliser operators describing a 4 qubit linear cluster state.
\\
\indent Extending this scheme to an $N$ qubit linear cluster state
is straightforward [Fig. \ref{fig:linear}].
Initially prepare $N$ qubits (for simplicity assume $N$ even) in the
state $|\phi\rangle_N = |00...00\rangle_N$. The stabiliser group for
this state is generated by $Z_j$, $j=[1,2,...,N]$. Hadamard
rotations are performed on all odd numbered qubits except for the
two at the centre of the chain and the following operators are
measured,
\begin{equation}
\begin{aligned}
&\text{Step 1} \quad \quad X_{N/2}X_{N/2+1}\\
&\text{Step 2} \quad \quad Z_{N/2-1}Z_{N/2} \quad \quad \text{and} \quad
X_{N/2+1}X_{N/2+2}\\
&\text{Step 3} \quad \quad X_{N/2-2}X_{N/2-1} \quad \text{and} \quad
Z_{N/2+2}Z_{N/2+3}\\
&\text{Step } N/2 \quad \quad Z_1Z_2 \quad \text{and} \quad X_{N-1}X_N.
\end{aligned}
\end{equation}
Hadamard gates are again applied to all odd numbered qubits, after
which the generators of the stabiliser set are identical to a linear
cluster state.  Although additional links in the cluster are created
sequentially, the first link is made from the centre of the chain
and subsequent links formed from this point, increasing the number
of possible operations that can be done in parallel. Using this
method, an $N$ qubit linear cluster state can be prepared using
$N/2$ time steps (for $N$ even) or $N/2+1$ time steps (for $N$ odd).
\\
\indent The preparation of linear cluster states via the QTP is
useful in the preparation of multi-qubit entangled systems, however
Nielsen \cite{Nielsen2} has shown that linear cluster states are
insufficient for universal quantum computation.  The original
proposal of Raussendorf and Briegel suggested a 2-D tiled cluster
state be used, however the QTP combined with single qubit gates is
insufficient to create such a state directly. Universality can be
achieved if we employ the results of Aliferis and Leung \cite{TQC1}
and their work into TQC. The QTP already assumes that single qubit
gates can be performed directly on data qubits, hence the ability to
simulate \emph{any} entangling gate between two data qubits is
sufficient for universality \cite{dodd}. We demonstrate explicitly
how a CZ gate can be simulated using the QTP and direct single qubit
operations.
\\
\indent Consider an arbitrary two qubit state $|\psi\rangle_{12} =
\alpha|00\rangle+\beta|01\rangle+\gamma|10\rangle+\delta|11\rangle$
and a third data qubit prepared in a
$|+\rangle_3=(|0\rangle_3+|1\rangle_3)/\sqrt{2}$ state that acts as
an ancilla.  Using the QTP a $Z_1Z_3$ operator measurement is
performed on the control and ancilla qubit (again we assume that the
qutrit returns to Alice and $+1$ eigenstates are projected, if not
the classical measurement record can be used to correct the state
using $X$ and/or $Z$ gates). This operation takes the combined
qubit/ancilla state to $|\psi^{'}\rangle = |\psi\rangle \otimes
|+\rangle  \rightarrow
\alpha|000\rangle+\beta|010\rangle-\gamma|101\rangle-\delta|111\rangle$.
A Hadamard gate is applied to the target qubit taking the combined
qubit/ancilla state to $|\psi^{'}\rangle =
\alpha|0+0\rangle+\beta|0-0\rangle-\gamma|1+1\rangle-\delta|1-1\rangle$.
An $X_2X_3$ operator measurement is now performed on the ancilla and
target qubits taking the state to $|\psi^{'}\rangle =
\alpha|0++\rangle
-\beta|0--\rangle-\gamma|1++\rangle-\delta|1--\rangle$. Performing a
Hadamard rotation on the target qubit leads to
\begin{equation}
\begin{aligned}
|\psi^{'}\rangle =
&(\alpha|00\rangle-\beta|01\rangle-\gamma|10\rangle-\delta|11\rangle)_{12}\otimes |0\rangle_3 + \\
&(\alpha|00\rangle+\beta|01\rangle-\gamma|10\rangle+\delta|11\rangle)_{12}\otimes |1\rangle_3.
\end{aligned}
\end{equation}
The ancilla qubit is now measured in the computational basis.  If it is measured in the $|0\rangle$ state, local phase gates
are applied to both the control and target qubit.  If the ancilla is measured in the $|1\rangle$ state, a local phase gate is
applied to the control qubit.  After these corrections the state has been transformed from
$|\psi\rangle$ to $\text{CZ}|\psi\rangle$.  Therefore, using specific operator measurements and local gates,
a CZ gate can be effectively simulated across two qubits by introducing a third ancilla.
\\
\indent Since a CZ gate can be directly implemented in this scheme,
and we have assumed that single qubit operations can be implemented
directly on data qubits, universal computation is possible using the
QTP. The simulation of CZ gates also allows for the preparation of
arbitrary cluster states (if desired) using the standard method of
linking un-entangled $|+\rangle$ states with CZ gates.
\\
\section{Conclusions}

\indent We have presented an information-free quantum bus, based on
qutrits that acts to mediate entanglement between data qubits
pairwise. To clarify this protocol, we have explicitly shown an
implentation that uses spatial adiabatic passage with a spatially
defined qutrit.  Our protocol allows for deterministic $XX$ and $ZZ$
operator measurements to be performed on separate data qubits. We
have demonstrated how this restricted set of operator measurements,
combined with the ability to do single qubit operations directly on
data qubits, allows for the preparation of $N$ qubit linear cluster
states and simulation of controlled phase (CZ) gates between two
data qubits. This approach to direct synthesis of operator
measurements may have significant application to improving the
efficiency of quantum operations, and constitutes a different
approach to the generation of remote entanglement from flying qubit
methods.
\\
\indent The authors thank J.Cole, A.Fowler, and K. Maruyama for
helpful discussions.  This work was supported by the Australian
Research Council, US National Security Agency (NSA), Advanced
Research and Development Activity (ARDA) and Army Research Office
(ARO) under contract W911NF-04-1-0290.

\end{document}